\documentclass[preprint,amsmath,amssymb,aps,doublespace]{revtex4}
\usepackage[dvips]{graphicx}
\usepackage{setspace}
\usepackage{amsmath}
\usepackage{amsfonts}
\usepackage{amssymb,color}
\usepackage{subfigure}
\usepackage{graphicx,color}
\usepackage{ifpdf}
\usepackage[latin1]{inputenc}

\begin{document}

\title{Population dynamics in the triplet annihilation model with a mutating reproduction rate}

\author{
Ronald Dickman,\footnote{email: dickman@fisica.ufmg.br}
}

\address{
Departamento de F\'{\i}sica and
National Institute of Science and Technology for Complex Systems,\\
ICEx, Universidade Federal de Minas Gerais, \\
C. P. 702, 30123-970 Belo Horizonte, Minas Gerais - Brazil
}

\date{\today}

\begin{abstract}
I study a population model in which the reproduction rate $\lambda$ is
inherited with mutation, favoring fast reproducers in the short term, but conflicting
with a process that eliminates agglomerations of individuals. 
The model is a variant of the {\it triplet annihilation model} introduced
several decades ago [R. Dickman, Phys. Rev. B~{\bf 40}, 7005 (1989)] in which organisms (``particles'')
reproduce and diffuse on a lattice, subject to annihilation when
(and only when) occupying three consecutive sites. For diffusion rates below a 
certain value, the population possesses two ``survival strategies'': 
(i) rare reproduction ($0 < \lambda < \lambda_{c,1}$), in which a low density 
of diffusing particles renders triplets exceedingly rare, and 
(ii) frequent reproduction ($\lambda > \lambda_{c,2}$). 
For $\lambda$ between $\lambda_{c,1}$ and $\lambda_{c,2}$
there is no active steady state.  In the rare-reproduction regime, 
a mutating $\lambda$ leads to stochastic boom-and-bust cycles in which
the reproduction rate fluctuates upward in certain regions, only to
lead to extinction as the local value of $\lambda$ becomes excessive.
The global population can nevertheless survive due to the presence 
of other regions, with reproduction rates that have yet to drift
upward.
\end{abstract}

\maketitle

\section{Introduction}

My regrettably few  encounters with Dietrich Stauffer left an important imprint.  
He always impressed with his unconventional, even provocative attitude, his honesty, generosity,
and sense of humor.  I first met Prof. Dr. Stauffer at a Rutgers Meeting
in the early '80s.  I'd been enumerating lattice animals as part of my PhD
project on nucleation in lattice gases.  When he asked what computer
I used for the task, I simply gestured to indicate that this was hand enumeration.
But a year later, having exhausted the possibilities of this approach (and/or my 
patience with filling notebooks with strange little diagrams),
Stauffer's question led me to develop Monte Carlo methods for
counting lattice animals\cite{animals}.

Toward the end of the 1980s, on a visit to Germany at the kind invitation 
of Prof. Kurt Binder, I met Stauffer at his lab in J\"ulich, and described my ongoing study 
of the triplet annihilation model, the subject of the present work.  Hearing my explanation 
that particles could only be annihilated in groups of three,
he turned and said, ``Oh, you're studying guerrilla warriors who 
spread out to evade detection from the air.''  This colorful and somewhat poignant interpretation stuck with me, 
as did his advice (in reply to a letter sent shortly before defending my PhD, in which I confessed to having 
become ``addicted'' to Monte Carlo
simulation), that I simply ``enjoy my addiction.''  And so I have.
In a handful of encounters in Europe and Brazil, we chatted about
science, politics and other matters. (I'm afraid my answers were sometimes 
disappointing: Seen any good movies recently? - My tastes were too art-house. 
Your favorite  football team?  - I had none.) 

Dietrich Stauffer made major contributions to percolation theory, so it is
fitting that the triplet annihilation model studied here belongs to the (rather broad) class 
of models with a
{\it directed} percolation (DP)-like phase transition. In the stochastic particle systems of interest here, the 
directed spatial axis of DP corresponds to time;
the (continuous) phase transition is between an active and an absorbing state.  The simplest example is the {\it contact process}
(CP) \cite{harris74,liggett,marro,odor07,henkel},
in which each site of a lattice is either vacant or occupied by a single
particle.  The latter die at unit rate, and attempt to reproduce at rate
$\lambda$, choosing a neighboring site at random and adding a particle
if the chosen neighbor is vacant.  In the infinite-size limit there is a
transition between an active phase (for $\lambda > \lambda_c$) with a nonzero particle density and an absorbing phase (all sites vacant).  To the CP dynamics, the triplet annihilation model \cite{tam89,tam90} adds diffusion (nearest-neighbor hopping) and a modified death process, in which only groups of three particles are
removed.   The result is that (for not too rapid diffusion) an active phase exists for small $\lambda$ (with a very low abundance of triplets) as well as for large $\lambda$, but not for intermediate values.
Here, I explore the effect of a reproduction rate,
$\lambda_i$ for each particle $i$, that is inherited from its parent
with some random variation.  Depending on the degree of variability, this can lead to a population with minimal reproduction,  
to extinction, or  to long-time survival punctuated by a series of local population
explosions and collapses. 

The remainder of this paper is organized as follows. In the next section I define the model and describe the simulation algorithm.  Sec. III reports simulation results, followed by a summary and discussion in Sec. IV.

\section{Model}

\subsection{Triplet annihilation model}

The triplet annihilation model (TAM) \cite{tam89,tam90} is a stochastic particle
system on a lattice, a ring of $L$ sites in the present study \cite{dimnote}. 
Each site can be either vacant or occupied by a (single) particle.  Transitions between configurations occur through three basic processes:

 \begin{itemize}
 \item Creation: Each particle attempts independently, at rate $2\lambda$, to create a new particle at its nearest-neighbor to the left or right.  The
 attempt is successful if the chosen neighbor is unoccupied.
 
 \item Hopping: Each particle attempts to hop, at rate $2d$, to its nearest-neighbor to the left or right.  The
attempt is successful if the chosen neighbor is unoccupied. 
 
 \item Annihilation: If sites $i-1$, $i$ and $i+1$ are all occupied, we say
 there is a {\it triplet} centered at site $i$.  Triplets are annihilated a rate of unity.  (All three particles are removed in this event.) 
 \end{itemize}

\noindent From this definition it is evident that the configuration with all sites
vacant is absorbing.  (The parameterization of the rates used
here differs from that used in the earlier studies, although it has no qualitative effect on the phase diagram.) 

Note that hopping tends to destroy triplets, whereas creation tends to
generate them, raising the possibility of subtle interactions between
these processes.
The original motivation for studying the model was to investigate
what effect such conflicting tendencies might have on the phase diagram and critical behavior.\cite{aside}

\begin{figure}[h!]
\centering
  \includegraphics[scale=0.6]{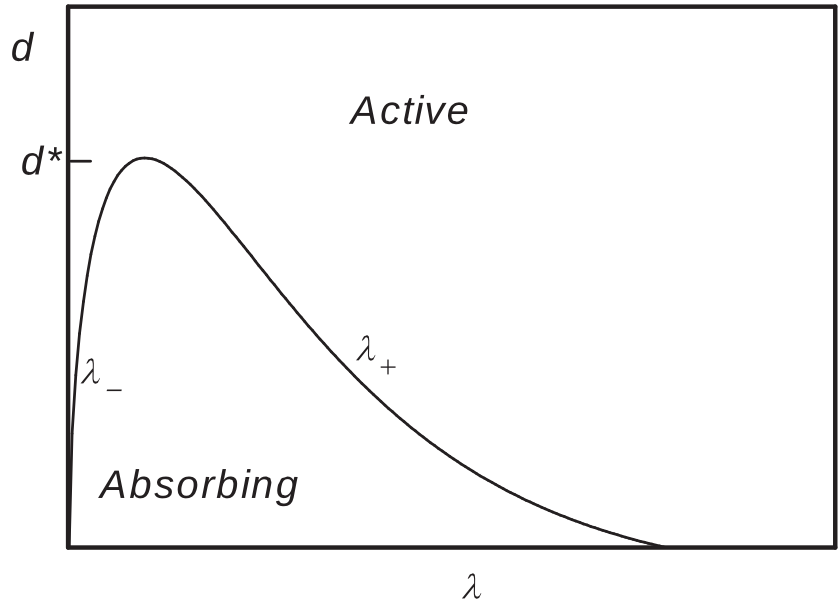}
  \caption{Schematic of the triplet annihilation model phase diagram in the reproduction rate-hopping rate ($\lambda - d$) plane.  The phase boundary is a line of continuous phase transitions between the active and absorbing phases.  For hopping rates $d > d^*$, there is activity for {\it any} $\lambda > 0$, while for $d < d^*$, the active phase exists for large $\lambda$, but also for very small $\lambda$. The
present study focuses on the region bounded by $\lambda = 0$
and the small-$\lambda$ branch of the phase boundary, $\lambda_- (d)$.
}
  \label{PD}
\end{figure}

\begin{figure}[h!]
\centering
  \includegraphics[scale=0.6]{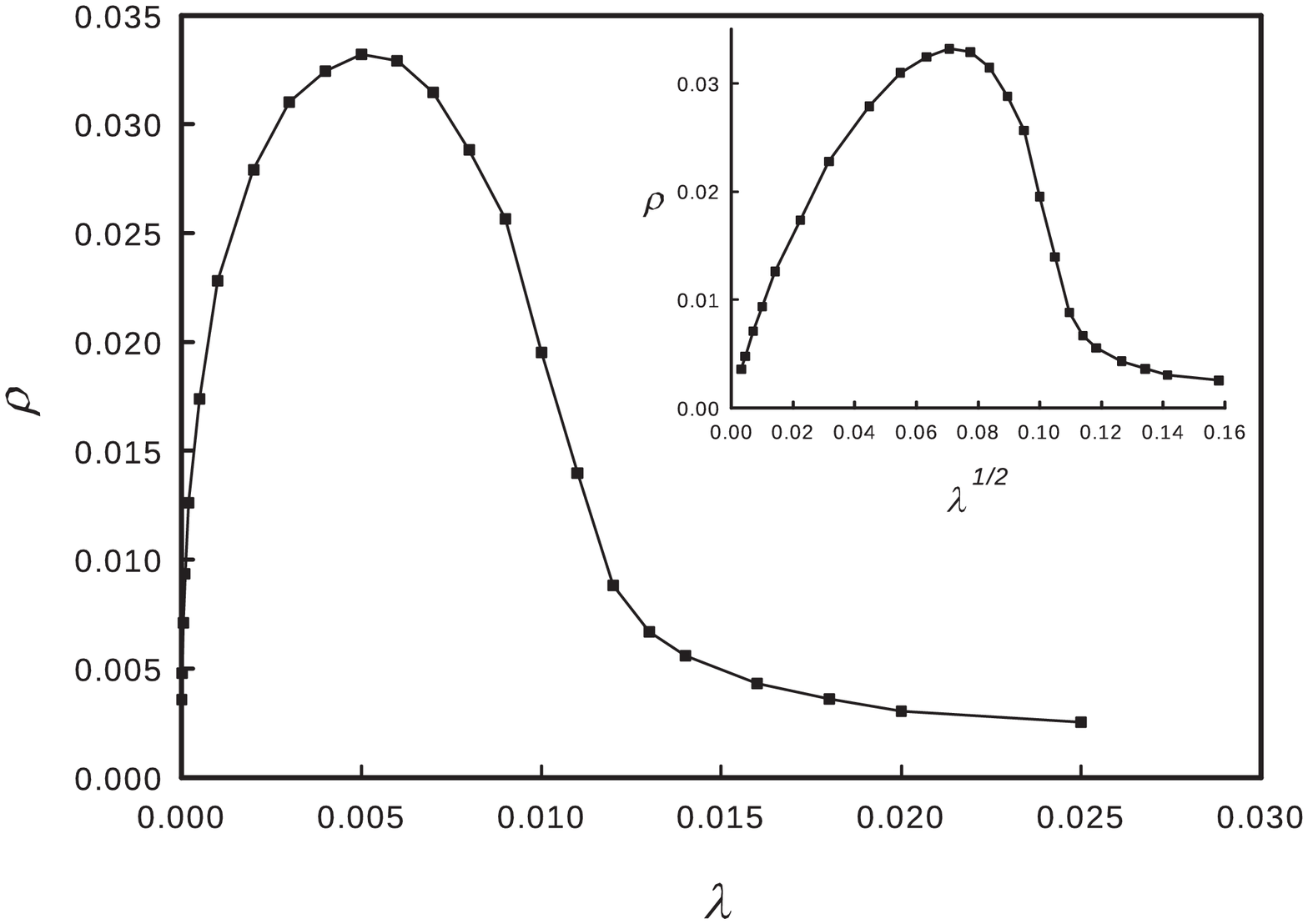}
  \caption{Quasistationary particle density $\rho$ versus reproduction rate
 $\lambda $ in the TAM, for hopping parameter $d = 1.2346$,
 system size $L = 10^4$.  Error bars are smaller than the symbols.  
 Inset: the same data plotted versus $\sqrt{\lambda}$.
}
\label{RvLamb}
\end{figure}

The phase diagram, shown schematically in Fig.~\ref{PD} is indeed
intriguing.  In simpler lattice birth-and-death models such as the contact process, sustained activity (i.e., a nonzero particle density, $\rho > 0$), requires a reproduction rate greater than
some value $\lambda_c > 0$.  The presence, in the TAM, of activity for
arbitrarily small $\lambda$ is due, of course, to the death rate being
proportional to the density of {\it triplets}, $\rho_3$, rather than of particles, $\rho$.
In the presence of hopping $(d > 0)$, there is an interval 
$(0, \lambda_-(d))$ of
reproduction rates such that $\rho_3 <\!\! < \rho$, permitting 
survival of activity.
Of particular interest in the present work is the portion of the active
phase bounded by $\lambda = 0$ 
and the small-$\lambda$ branch of the phase boundary, $\lambda_- (d)$,
because in this region, paradoxically, increasing the reproduction rate
leads to extinction, due to the increased abundance of triplets.
An example of this behavior is shown in Fig.~\ref{RvLamb}.  The
quasistationary particle density  \cite{QS} peaks near $\lambda \simeq 0.005$
and falls rapidly for higher values.  For this hopping rate ($d = 1.2346$), the upper critical
reproduction rate, $\lambda_+ \simeq 1.17$, much larger than the values
considered here;  preliminary studies yield $\lambda_- \simeq 0.03$.

Although critical behavior is not the principal subject of the present study, it is worth
noting that the entire phase boundary appears to be a line of
critical points in the directed percolation universality class, and that
the particle density $\rho \sim \lambda^{1/2} $ as $\lambda \to 0$.
This property, as well as the qualitative form of the phase
diagram, is captured by mean-field theory at the pair level \cite{tam89,tam90}.
Several works have investigated the phase diagram and closely related 
models.
Katori and Konno \cite{katori} established some general properties of systems
like the TAM, while Poland \cite{poland} used time-series expansions to
investigate population dynamics in the TAM and related models.
A model with annihilation of quadruplets was analyzed by \'Odor \cite{odor2004,odor2004a}.

\subsection{TAM with mutating reproduction rate (MTAM)}

In the present study, the reproduction rate is no longer a fixed parameter.  Instead, each particle $i$ is assigned a value $\lambda_i$ when created, and carries this value during its lifetime, as it hops on the lattice.  If particle $i$ creates a new particle $j$, then $\lambda_j$ is
inherited from the parent with random variation:

\begin{equation}
\lambda_j = \max \left[ \lambda_i \!+\! \Delta, \, \lambda_{min} \right],
\label{inherit}
\end{equation}

\noindent where $\Delta$ is a random variable with zero mean, 
and $\lambda_{min}$ is a small basal reproduction rate.
I consider both additive and multiplicative schemes.  In the former,
$\Delta$ is simply chosen from a distribution uniform on 
$[-\gamma, \gamma]$.  In the multiplicative case,
$\Delta = \xi \lambda_i$, where $\xi$ is uniform on 
$[-\gamma, \gamma]$.  Each realization of the process begins with
all sites occupied, and $\lambda_i = \lambda_0, \, \forall i=1,...,L$.
In what follows I refer to this as the mutating triplet annihilation model (MTAM).

\subsection{Simulation algorithm}

I developed an event-driven (rejection-free) Monte Carlo simulation algorithm for the present study.
The basic requirements this type of simulation are lists of the possible transitions and pointers
to facilitate updating the lists.  In the TAM simulation one must maintain a list of triplets and
another list of nearest-neighbor particle-hole pairs (PHPs).  One maintains, in addition, vectors carrying (i) the
occupation status of each site; (ii) whether sites $i$ and $i+1$ belong to a PHP, and if so, the position on
the PHP list, and which of the two sites is occupied; (iii) whether site $i$ is the central site of a triplet, and if
so, the corresponding position on the triplet list.  In the case of mutating reproduction rates, a fourth vector
carries $\lambda_i$ associated with each occupied site $i$.

Given the numbers, $N_{trp}$ and $N_{php}$ of triplets and PHPs, respectively,
the total transition rate out of the current configuration is,

\begin{equation}
R = N_{trp} + (d + \lambda) N_{php} \equiv R_{ann} + R_{hop} + R_{rep},
\label{defR}
\end{equation}

\noindent where in the second equality we define the total rates of annihilation, hopping and reproduction.
The next transition is taken to be of type $j$ (annihilation, hopping or reproduction),
with probability $P_j = R_j/R$.  Then the specific event on the corresponding list is selected
(with uniform probability over the list), the new configuration constructed, the lists updated, and
the time advanced so: $t \to t + 1/R$.

In the case of mutating reproduction rates,  i.e., the MTAM, we must alter the expression for 
the total creation rate to $R_{rep} = \Lambda$, where 

\begin{equation}
\Lambda = \sum_{i=1}^{N_{php}} \lambda_{j(i)},
\label{DefLambda}
\end{equation}

\noindent where the sum is over PHPs and $j(i)$ denotes the occupied site in the $i$th PHP.  In this case, of course, the PHP for creation must be chosen with a probability proportional to the associated $\lambda$ value.  This is done by letting $\tilde{\Lambda} = \Lambda z$,
where $z$ is a random number uniform on $[0,1]$, and then evaluating the partial sums $\Lambda_k$, with 
$\Lambda_1 = \lambda_{j(1)}$ and,
\begin{equation}
\Lambda_k = \Lambda_{k-1} + \lambda_{j(k)},
\label{DefLambdak}
\end{equation}

\noindent until $\Lambda_k > \tilde{\Lambda}$, at which point creation is realized in the $k$th PHP.  While more efficient methods
for choosing the PHP exist, this approach is simple and unlikely to increase run times significantly since reproduction counts for
only a small fraction of events and the number of PHPs is not very large in the regime of interest.

\section{Results}

Given the sensitivity of the TAM to small changes in the reproduction rate, and the counterintuitive nature of the
response, it is of interest to examine how a population evolves under a mutating rate.

In this section I report preliminary results for the time evolution and spatial structure of the MTAM.
The fixed simulation parameters are $d = 1.2346$, initial creation rate $\lambda_0 = 0.005$, basal creation rate $\lambda_{min} = 0.001$,
and system size $L = 2\,000$.  Studies run to a maximum time of $10^6$ unless the absorbing state is attained prior
to this time.  (For these parameters, but without mutation, the probability of a realization surviving to time $10^6$
is about 0.86, and the quasistationary particle density is $\rho = 0.0331(1)$.)

\begin{figure}[h!]
\centering
  \includegraphics[scale=0.55]{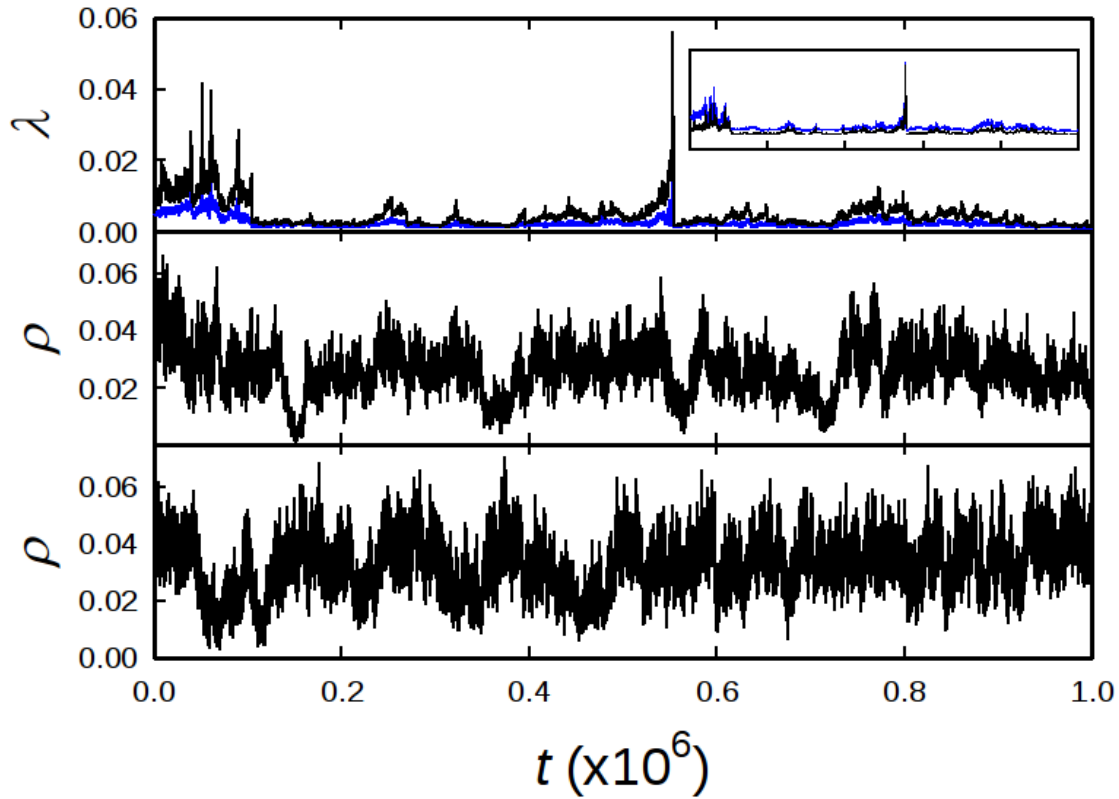}
  \caption{Time series of the particle density $\rho(t)$ and reproduction rate in the TAM and MTAM with hopping parameter $d = 1.2346$,
and  system size $L = 2000$.  Lower panel: $\lambda = 0.005$, without mutation (TAM). 
Middle panel:  $\rho(t)$ for the MTAM with mutation parameter $\gamma = 0.1$
(multiplicative case).  Upper panel: $\bar{\lambda}(t)$(lower) and $\lambda_{max}(t)$ (upper) in the same realization as the middle panel.
Inset: comparison of time series for $\bar{\lambda}(t)$ (upper curve) and the standard deviation of the set of $\lambda_i$ values at time $t$ (lower).
}
\label{ts1}
\end{figure}

\begin{figure}[h!]
\centering
  \includegraphics[scale=0.45]{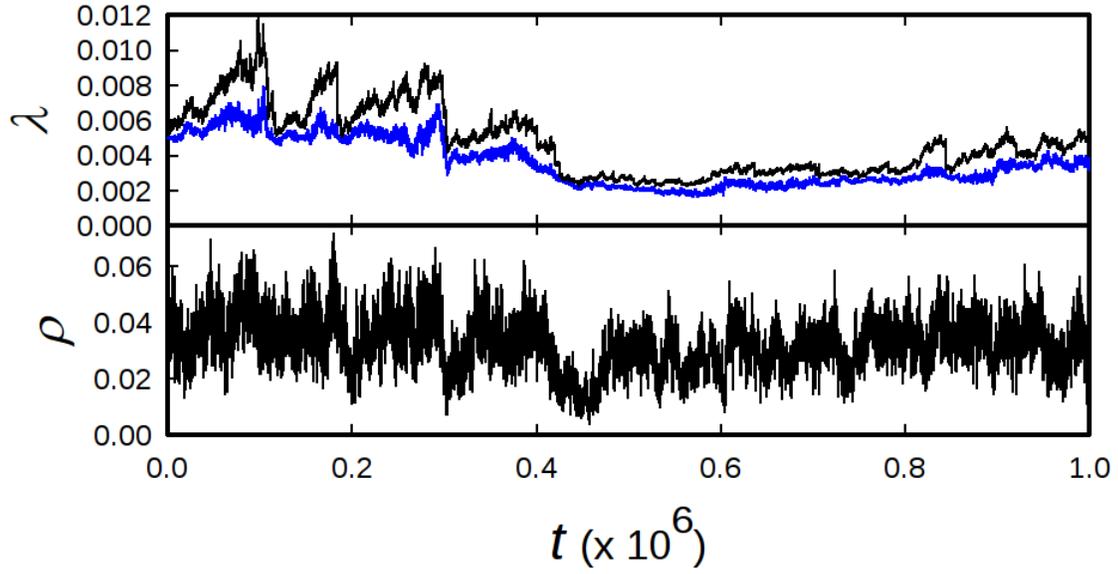}
  \caption{MTAM: Time series of the particle density $\rho(t)$ and reproduction rate as in Fig.~\ref{ts1}, except 
  $\gamma = 0.025$.
}
\label{ts3}

\end{figure}

\begin{figure}[h!]
\centering
  \includegraphics[scale=0.45]{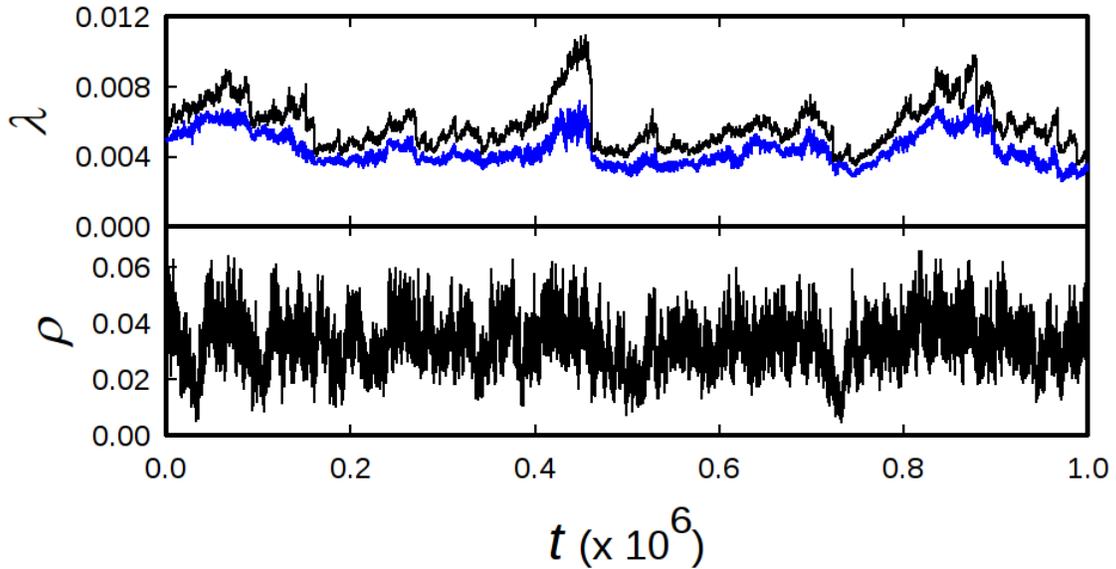}
  \caption{MTAM: Time series of the particle density $\rho(t)$ and reproduction rate, parameters as in Fig.~\ref{ts3}.
}
\label{ts4}
\end{figure}

To begin I consider the effect of multiplicative noise in the inherited reproduction rate, varying the noise
intensity $\gamma$ between $0.005$ and 0.25.   Most realizations survive until the maximum time:  For $\gamma = 0.025$,
for example, the survival probability is about 0.87.  Time series for the particle density show significant fluctuations, with or without mutation,
as shown in Fig.~\ref{ts1}.  This figure also shows the time series for $\bar{\lambda}(t)$, the instantaneous average of $\lambda$ over
all particles, as well as the instantaneous maximum, $\lambda_{max}(t)$.  These quantities exhibit a pattern of slow growth followed by a sudden crash.  The mean reproduction rate remains ``pinned'' near its basal rate of $10^{-3}$ for long periods.
The inset shows that the variations in the standard deviation of $\lambda$ mirror those of $\bar{\lambda}(t)$.
The particle density also exhibits crashes (followed by recovery, at least in this example), but the
relation between crashes in population and in $\lambda$ is unclear, since one may occur without the other.  
Similar examples of time series with
mutating reproduction rates are shown in 
Figs.~\ref{ts3} and \ref{ts4}.

\begin{figure}[h!]
\centering
  \includegraphics[scale=0.55]{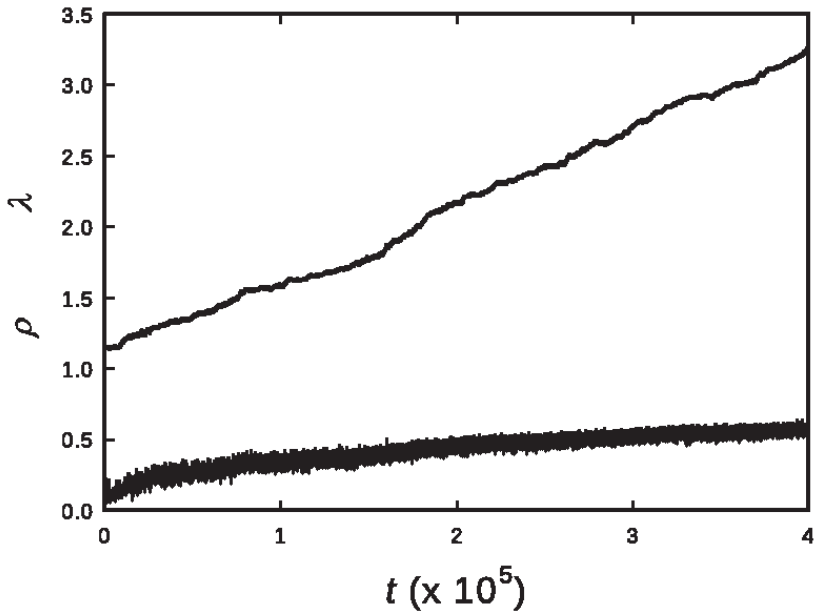}
  \caption{ MTAM: Time series of the particle density $\rho(t)$ and mean reproduction rate in with multiplicative noise,
  hopping parameter $d = 1.2346$,
initial reproduction rate $\lambda_0 = 1.15$, $\gamma = 5 \times 10^{-4}$, and
system size $L = 2000$.
} 
\label{runaway}
\end{figure}

For an initial reproduction rate $\lambda_0$ close to the {\it upper} phase boundary, $\lambda_+ (d)$,  
mutations lead, as expected, to steady growth in $\bar{\lambda}$ and the particle density.  An example is shown in Fig.~\ref{runaway},
for $\lambda_0 = 1.15$, close to $\lambda_+ = 1.17$.

\begin{figure}[h!]
\centering
  \includegraphics[scale=0.45]{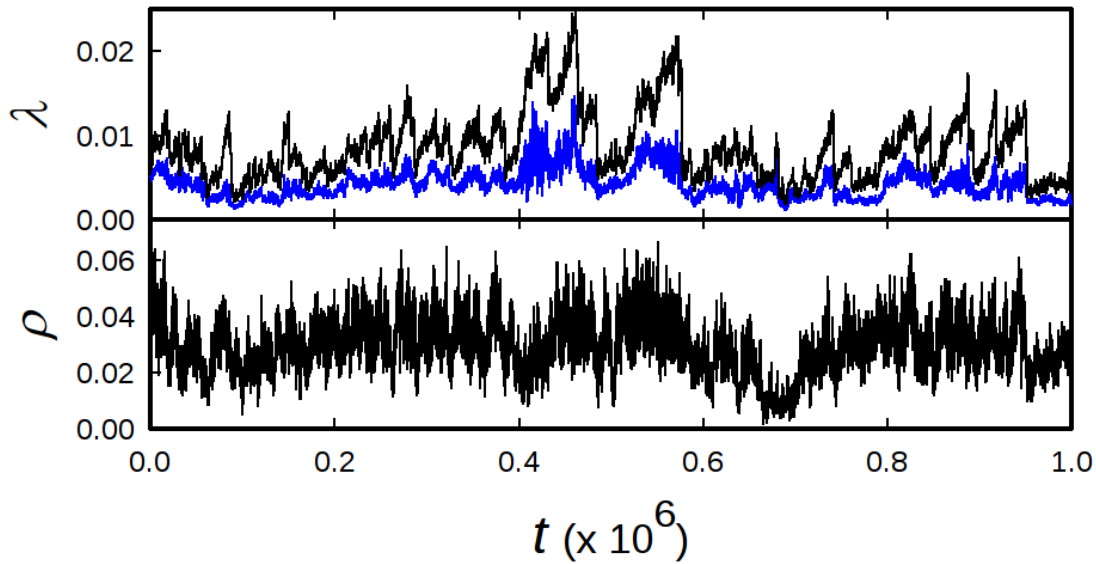}
  \caption{MTAM: Time series of the particle density $\rho(t)$ and reproduction rate, but with additive noise of intersity
  $\gamma = 5 \times 10^{-4}$.  Other parameters as in Fig.~\ref{ts1}.
}
\label{ts5}
\end{figure}

Time series for {\it additive} noise are generally similar to those for the multiplicative case, if we use
noise intensities $\gamma$ about two orders of magnitude smaller.  (Recall that in the multiplicative case,
the noise term is scaled by $\lambda \sim 0.005$.)  A typical example, for $\gamma = 5 \times 10^{-4}$,
is shown in Fig.~\ref{ts5}.  ``Pinning'' of $\bar{\lambda}$ near $\lambda_{min}$ is not observed under additive noise,
since the magnitude of mutations (i.e., of $\Delta \lambda$) is independent of $\lambda_i$. 
For $\gamma \geq 10^{-3}$ most realizations with additive noise do not survive until time $10^6$.

\begin{figure}[h!]
\centering
  \includegraphics[scale=0.45]{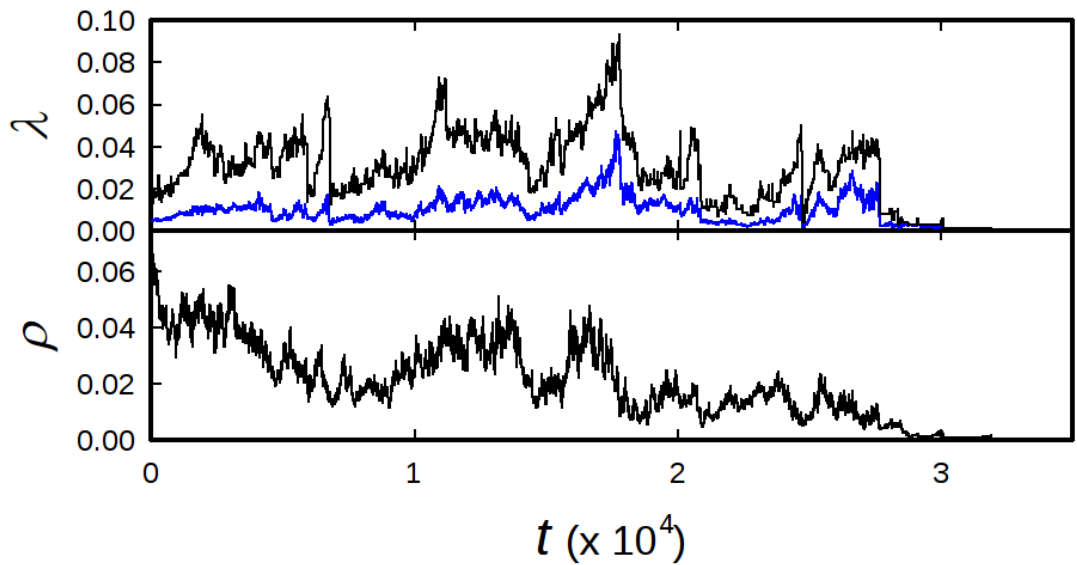}
  \caption{MTAM: Time series of the particle density $\rho(t)$ and reproduction rate, but with additive noise of intersity
  $\gamma = 0.005$.  Other parameters as in Fig.~\ref{ts1}.  The population goes extinct at $t = 3.192 \times 10^4$.
}
\label{ts6}
\end{figure}

An example of a population that goes extinct rapidly is shown in Fig.~\ref{ts6}.  Inspection of the time
series shows that even the exceptionally high values of $\bar{\lambda}$ for $t \simeq 1.7 \times 10^4$ do not lead
to an immediate population crash.  Partial extinction events near $t =1.8$ and $2.1 \times 10^4$ appear to cull
the rapidly reproducing subpopulations.  The population and reproduction rate continue to fluctuate, until,
at around $t = 2.8 \times 10^4$, the system attains a regime with small $\rho$ {\it and} small $\bar{\lambda}$.  This
combination makes appearance of individuals with  larger $\lambda$ values unlikely, and the population
goes extinct some time later.  This example shows that although high reproduction rates play a role in extinction, they may not
the direct or immediately  cause.  Further analyses will be required to assess the generality of this scenario.

\subsection{Spatiotemporal evolution}

\begin{figure}[h!]
\centering
  \includegraphics[scale=0.45]{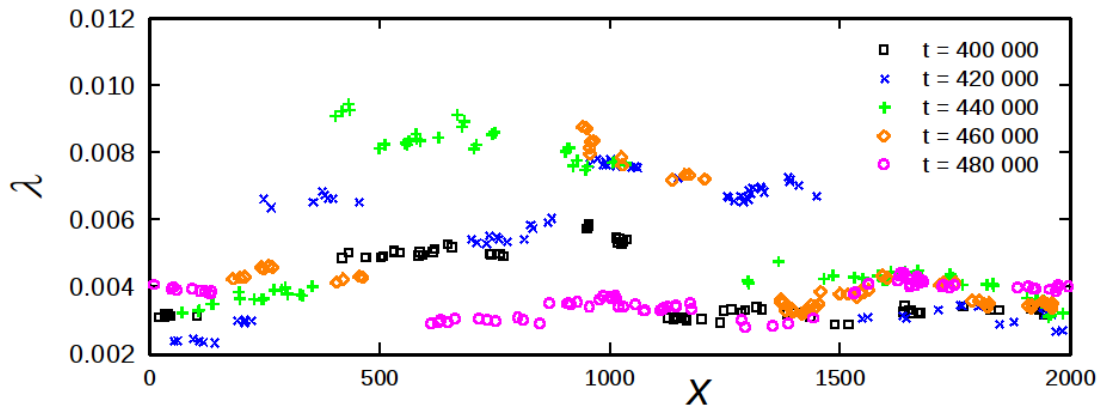}
  \caption{MTAM: Spatiotemporal evolution during a period of population growth and decay.  Each type of symbol (see
  legend) corresponds to a specific time in the time series of Fig.~\ref{ts4}.  Each symbol represents the 
  position ($x$) and reproduction rate ($\lambda$) of a single particle.
  }
\label{almbxhist1}
\end{figure}

An interesting aspect of the population dynamics is the appearance of localized subpopulations with a more or
less uniform reproduction rate.  The average reproduction rate within such a group may fluctuate upward until
it exceeds  a ``safe'' value, and the group dies out due to overproduction or triplets.  For parameter sets allowing
long-time survival, it appears that when such a group goes extinct, there remain several other subpopulations,
typically with small reproduction rates.  An example of such an evolution is shown in Fig.~\ref{almbxhist1},
which covers the period $t = 4.0 - 4.8 \times 10^5$ shown in Fig.~\ref{ts4}, during which the population grows
and then crashes.  Localized groups with restricted variations in $\lambda$ are evident; one does not find 
individuals with markedly different $\lambda$ values in the same region at the same time.  In the region
extending from $x \simeq 500 - 1000$, the mean value of $\lambda$ initially drifts upward. The region then becomes nearly
empty, to be reoccupied at the final time shown, presumably by particles that have diffused in from
nearby groups.  By contrast, the region between $x = 1500$ and 200 (i.e., 2200 under the periodic boundary) 
exhibits only modest variations in density and reproduction rate.  One may conjecture that
this dynamic homogeneity is what allows for long-term survival in a population undergoing
``boom-and-bust'' cycles in localized regions.  (Survival, one might say, lies in diversity.) Further studies, including the effect of varying system
sizes, will be needed to test these ideas.

\section{Summary and discussion}

I study the population dynamics of the triplet annihilation model (TAM) in a regime in which only a narrow range 
of (quite small) reproduction rates assure survival.  The the mutating version, MTAM, individual reproduction rates
are inherited with random mutation, leading to complex spatiotemporal dynamics.  Much
remains to be analyzed following this preliminary study, including analytic approximations such as deterministic and/or stochastic
partial differential equations.

There are further variations to be considered.  In the present context, for example, individuals with an
optimal reproduction rate (in the sense of Fig.~\ref{RvLamb}) may be seen as cooperators, and rapid reproducers
as cheaters or betrayers.  This raises the possibility of introducing mechanisms that foster cooperation in an evolutionary-game
context.  

Stochastic particle systems offer fascinating challenges for physicists and mathematicians.  Many
of their interpretations, as models of populations, epidemics, social systems, or even guerrilla warriors, are imperfect, serving
nevertheless as intuitively attractive metaphors for processes in the real world.  I wonder what fanciful interpretation
Dietrich Stauffer would have offered for the model considered here.

\newpage
\noindent {\bf Acknowledgment}

\noindent This work was supported by CNPq, Brazil, under grant No. 303766/2016-6.


\begin{thebibliography}{99}
\bibitem{dimnote}
             Thus the present study, like those of Refs. \cite{tam89}, \cite{tam90}, and \cite{poland}, is
             restricted to one dimension.  The behavior of the model in higher
             dimensions, or on fractal substrates, remains as an interesting topic
             for future study.
\end{thebibliography}
\end{document}